\documentclass[rmp, twocolumn, groupedaddress, superscriptaddress, showkeys]{revtex4}
\usepackage{latexsym}
\usepackage{graphicx}

\begin{document}
\title{Reverse Hall-Petch effect in ultra nanocrystalline diamond}

\author{Ioannis N. Remediakis}
\thanks{To whom correspondence should be addressed}
\email{remed@materials.uoc.gr}
\affiliation{Department of Materials Science and Technology, University of
Crete, 71003 Heraklion, Greece}
\author{Georgios Kopidakis}
\affiliation{Department of Materials Science and Technology, University of
Crete, 71003 Heraklion, Greece}
\author{Pantelis C. Kelires}
\affiliation{Department of Physics, University of Crete, 71003 Heraklion,
Greece}
\affiliation{Department of Mechanical Engineering and Materials Science and
Engineering, Cyprus University of Technology, 3603 Limassol, Cyprus}

\begin{abstract}
We present atomistic simulations for the mechanical response of ultra
nanocrystalline diamond, a polycrystalline form of diamond with grain
diameters of the order of a few nm. We consider fully three-dimensional model
structures, having several grains of random sizes and orientations, and employ
state-of-the-art Monte Carlo simulations. We calculate structural properties,
elastic constants and the hardness of the material; our results compare well
with experimental observations for this material. Moreover, we verify that
this material becomes softer at small grain sizes, in analogy to the observed
reversal of the Hall-Petch effect in various nanocrystalline metals. The
effect is attributed to the large concentration of grain boundary atoms at
smaller grain sizes. Our analysis yields scaling relations for the elastic
constants as a function of the average grain size.
\end{abstract}

\date{May 21, 2008}

\keywords{nanocrystalline materials; carbon; hardness \vspace{6mm} \\
Proceedings of the IUTAM Symposium on Modelling Nanomaterials and Nanosystems,
Aalborg, Denmark, May 19-22 2008; to be published in the IUTAM Bookseries by
Springer.
\vspace{6mm}}

\maketitle

\section{Introduction}

Most solids are polycrystalline, having grains in the micrometer to millimeter
range. As the percentage of atoms residing on grain boundaries is negligible,
the polycrystallicity only marginally affects the properties of the
material. In particular, mechanical properties of such solids usually depend
on bulk properties of the ideal material and the concentration of various
defects, such as cracks, dislocations, vacancies and interstitials; usually
the grain size is of minor importance. There are, however, properties where
grain size plays a key role, hardness being one of them.  The Hall-Petch law
states that the hardness ($H$) of polycrystalline metals increases with
decreasing average grain size ($d$), being a linear function of $d^{-n}$,
where $n>$0. The effect is attributed to the impedance of dislocation motion
due to grain boundaries (Bata and Pereloma 2004).

Modern nanotechnology makes it possible to synthesize nanocrystalline 
solids, i.e. polycrystalline solids with average grain sizes of a few nm. 
Such materials offer new possibilities for technological applications, 
mainly due to their unique mechanical properties (Meyers et al. 2006). The 
Hall-Petch law dictates that a nanocrystalline solid should have huge 
hardness, usually much higher than its usual polycrystalline phase. While 
this is true in most cases, Cu was found to become softer with decreasing 
grain size in the range between 3 and 7 nm (Schi{\o}tz et al. 1998). This 
was called the ``reverse Hall-Petch effect''. Later, it was found that many 
materials posses a ``strongest size'' (Yip 1998), which turned out to be 
around 15 nm for Cu (Schi{\o}tz and Jacobsen 2003). A similar effect has 
been observed recently for BN (Dubrovinskaia 2007).

The existence of a ``strongest size'' suggests that the mechanism of
undertaking mechanical load should be different in the nano-world. In this
region, the presence of dislocations no longer governs the mechanical response
on the material: as dislocations are extended defects, they cannot reside in
the limited space of nano-grains. Any external mechanical load will be
primarily undertaken by sliding along grain boundaries (Schi{\o}tz et
al. 1998, van Swygenhoven et al. 1999, Yamakov et al. 2004). This, in turn, is
a direct consequence of the enormous concentration of grain boundary atoms in
a nanocrystalline material. Imagine for example a cube of side $d$, containing
$N\times N \times N$ atoms. The fraction of surface atoms is roughly
proportional to 6$N^{2}$/$N^{3}$ = 6/$N$ $\approx $ 1 nm/$d$, ranging from 1
per million when $d$ is of the order of mm, to 30{\%} when $d$ is around 3 nm.

Contrary to a large number of studies for metals, very few studies have
addressed the mechanical properties of semiconductors and insulators as a
function of their grain size, although several pioneer workers have examined
the mechanical response of nanocrystalline ceramics (Demkowicz et al. 2007;
Szlufarska et al. 2005). Covalent solids are characterized by their open
structures and the strong directionality of their bonds. Such bonds should not
allow easy sliding along grain boundaries. At the same time, bonds on grain
boundaries will be considerably weaker than those in the bulk of grains, due
to the loss of the ideal local bonding geometry (Keblinski et
al. 1999). Recent studies in BN (Dubrovinskaia et al. 2007), together with the
well-established results for various metals, suggest that the effect might be
universal. To check whether this is the case, we study diamond, which
comprises the ideal test suite for this purpose. C atoms form perhaps the most
directional bonds known, as indicated by the supreme hardness and shear
modulus of diamond. In addition to being strong, C-C might in some cases break
and form a more stable structure, as threefold $sp^{2}$ C atoms are more
stable than fourfold $sp^{3}$ ones, the later being responsible for mechanical
failure in carbon-based materials (Fyta et al. 2006; Remediakis et al. 2007).

Ultra nanocrystalline diamond (UNCD) is a polycrystalline carbon-based 
material, with grain diameters mostly between 2 and 5 nm (Gruen 1999). It is 
a low-cost material with a potential for a wide range of applications due to 
its unique mechanical and electronic properties (Krauss et al. 2001). 
Despite the strong directional C-C bonds, resulting in inhomogeneity at the 
atomic scale, the material can be considered as isotropic at larger scales, 
as no particular orientation for the grain boundaries in UNCD seems to be 
favoured in the experiment (Gruen 1999) or simulations (Zapol et al. 2001, 
Kopidakis et al. 2007).

\section{Computational method}

We use fully three-dimensional atomistic models for UNCD, having grains of 
different random size and orientations separated by random grain boundaries. 
The simulations were performed using a continuous-space Monte Carlo method. 
We employ the many-body potential of Tersoff (Tersoff 1988), which provides 
a very good description of the structure and energetics for a wide range of 
carbon-based materials (Kelires 1994; Kelires 2000). This method, although 
considerably demanding computationally, allows for great statistical 
accuracy, as it is possible to have samples at full thermodynamic 
equilibrium. Such accuracy is necessary in order to capture all 
hybridizations of C. 

We model UNCD by a periodic repetition of cubic supercells that consist of 
eight different regions (grains). The number of grains in the unit cell 
guarantees the absence of artificial interactions between a grain and its 
periodic images. The grains have random shapes and sizes, and are filled 
with atoms in a randomly oriented diamond structure. The method we use is 
identical to the method used by Schi{\o}tz et al. (Schi{\o}tz et al. 1998; 
Schi{\o}tz and Jacobsen 2003). To achieve a fully equilibrated structure for 
each grain size, we perform four steps: first, the structure is compressed 
and equilibrated at constant volume at 300 K, in order to eliminate large 
void regions near some grain boundaries, that are an artifact of the 
randomly generated structure. In the second step, we anneal the system at 
800 K allowing volume relaxation and quench down to 300 K. Third, we anneal 
once more, at 1200 K this time, in order to ensure full equilibration. 
Fourth, we fully relax the structure at 300 K allowing for changes in both 
volume and shape of the unit cell.

\begin{table}
\caption{Properties of some characteristic UNCD samples at 300K: average grain
size ($d$, in nm); number of atoms in the simulation cell ($N$);
percentage of three-fold atoms in the cell ($N_{3}$, at {\%}); mass
density ($\rho$, in g/cc); cohesive energy ($E_{c}$, in eV
per atom); bulk ($B$), Young's ($E$) and shear ($G$)
moduli (all in GPa); Vickers hardness ($H$), as estimated from the
theory of Gao et al. (Gao et al. 2003). For comparison, the corresponding
values for single-crystal diamond, calculated with the same method (‰elires
1994), are shown in the last line.}

\begin{center}
\begin{tabular}{rrrrrrrrr}
\hline
$d$& 
$N$& 
$N_{3}$& 
$\rho $& 
$E_{c}$& 
$B$& 
$E$& 
$G$& 
$H$ \\
\hline
2.4& 
18,528& 
26& 
3.22& 
-7.06& 
323& 
808& 
373& 
89.8 \\
\hline
3.4& 
53,494& 
12& 
3.30& 
-7.10& 
363& 
939& 
439& 
90.9 \\
\hline
4.4& 
116,941& 
9& 
3.40& 
-7.15& 
384& 
987& 
461& 
91.6 \\
\hline
$\infty $& 
$\infty $& 
0& 
3.51& 
-7.33& 
443& 
1066& 
485& 
94.2 \\
\hline
\end{tabular}
\label{tab1}
\end{center}
\end{table}

\begin{figure}
\centerline{\includegraphics[width=\columnwidth]{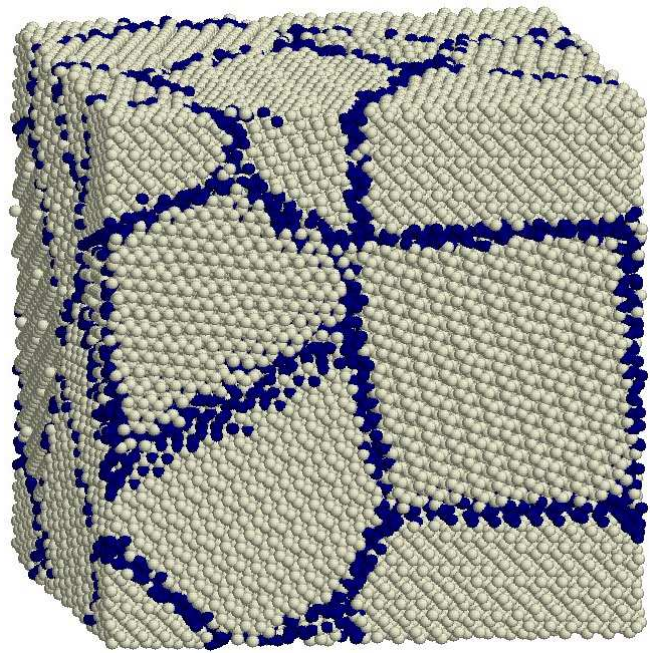}}
\label{fig1}
\caption{ Relaxed model structure for Ultra-Nanocrystalline Diamond, with an
average grain size of 4.4 nm. Atoms are coloured according to their number of
neighbours ($z$) and average angle between their bonds ($\theta $). Atoms in
the diamond structure ($z$=4, $\cos\theta $=-1/3) are coloured gray; all other
atoms are coloured blue.}
\end{figure}

\section{Structure and elastic moduli}

The relaxed structure for a typical sample is shown in Fig. 1. The grain 
boundaries are a few atomic diameters across, in accordance with experiments 
showing widths of 0.2-0.5 nm (Gruen 1999). Atoms at grain boundaries are 
either three-fold coordinated or form bonds at different lengths or angles 
from those observed in diamond. The structural and elastic properties for 
characteristic samples are summarized in Table I. The fraction of the 
three-fold atoms in the samples is about 1/10 for grain sizes between 3.5 
and 4.5 nm; in experiment, it was observed that the fraction of atoms 
residing at grain boundaries is close to 10{\%} for similar crystallite 
sizes (Gruen 1999). The density of UNCD increases with increasing 

grain size, as both the percentage of \textit{sp}$^{2}$ atoms and the 
concentration of voids is decreased. The cohesive energy decreases with 
increasing grain size, suggesting that most grain boundary atoms should be 
considered as defective ones.

The bulk modulus of UNCD decreases with decreasing grain size, despite the 
increase in the fraction of energetically favorable \textit{sp}$^{2}$ atoms. 
Although they have lower energy, \textit{sp}$^{2}$ C atoms are actually 
easier to deform compared to \textit{sp}$^{3}$ ones. This can be 
demonstrated by employing the concept of local bulk moduli (Kelires 2000). A 
similar analysis of our samples yields the average local bulk modulus of 
\textit{sp}$^{2 }$atoms to be around 250 GPa, while the average bulk modulus 
of \textit{sp}$^{3 }$atoms is around 420 GPa. This agrees very well with 
experimental observations for UNCD where the grain boundaries have been 
found to have much lower local bulk moduli than the bulk of grains (Pantea 
et al. 2006).

\begin{figure}
\centerline{\includegraphics[width=\columnwidth]{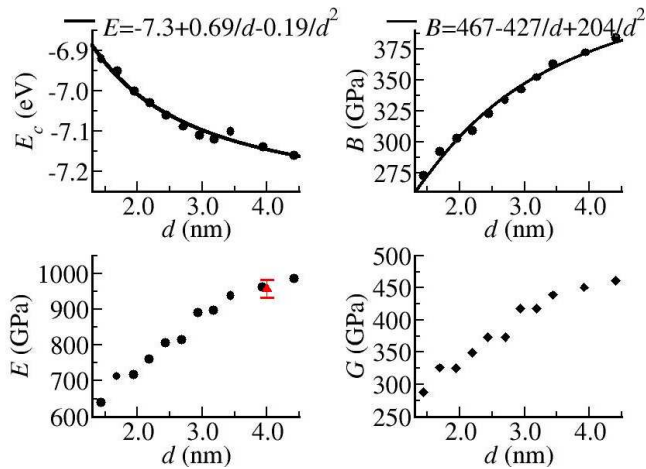}}
\label{fig2}
\caption{Cohesive energy ($E_{c})$ , bulk modulus ($B$), Young's modulus ($E$)
and shear modulus ($G$) of UNCD as a function of the average grain size. Solid
lines in the upper panels are fits to the simulations. The triangle in the
graph for $E$ represents an experimental measurement (Espinosa et al. 2006). }
\end{figure}

\section{Hardness}

Decrease of all elastic moduli with decreasing average grain size suggests 
that the hardness of the material should also drop with decreasing grain 
size, as the hardness of many materials is proportional to the Young's or 
shear modulus (Brazhkin et al. 2002); in particular, the hardness of all 
known carbon-based materials has been found to be between 10{\%} and 16{\%} 
of the Young's modulus (Robertson 2002). Elastic moduli are reliable probes 
of hardness for nanocrystalline solids, as the later cannot contain extended 
defects, such as cracks or dislocations, that have characteristic lengths 
exceeding the size of the grains. 

To get a quantitative description of hardness, we use the theory of Gao et 
al., who correlate the Vickers hardness of covalent crystals with the 
electron density per bond and the energy gap of the material (Gao et al. 
2003). The hardness of a complex material is the geometrical mean of the 
values of hardness for each subsystem. Here, we consider each individual 
pair of neighbouring C atoms as a subsystem. The density of valence 
electrons in a particular bond can be obtained from the bond length and the 
coordination numbers of the two atoms that participate in this bond. The 
calculated hardness of UNCD is shown in Fig. 3, demonstrating the existence 
of the reverse Hall-Petch effect for this material.

\begin{figure}
\centerline{\includegraphics[width=\columnwidth]{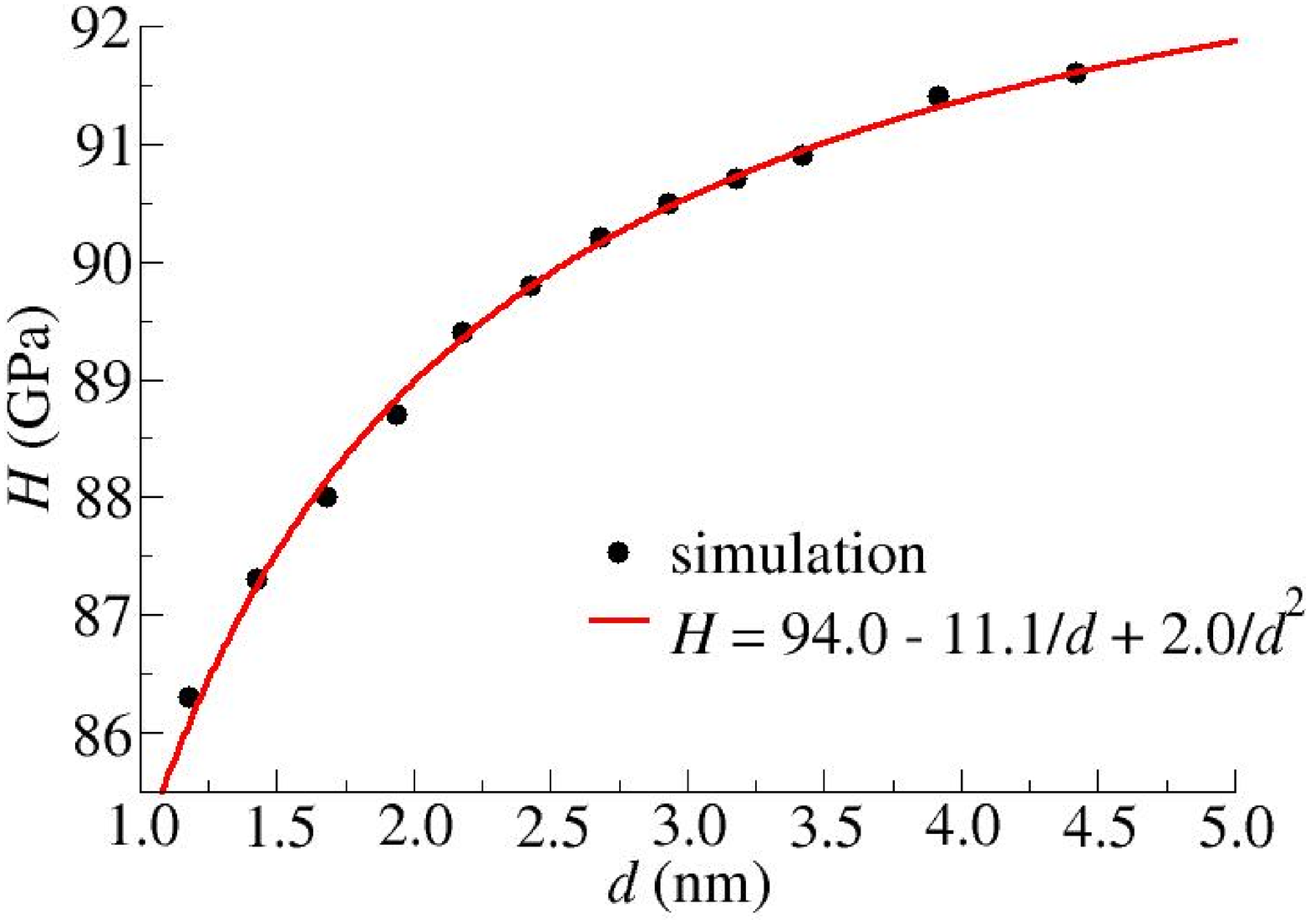}}
\label{fig3}
\caption{Estimated hardness of UNCD as a function of the average 
grain size. Hardness is calculated based on the theory of F. Gao et al. for 
covalent solids (Gao et al. 2003).}
\end{figure}

\section{Scaling laws for the properties of UNCD}

To understand the mechanism behind softening of UNCD, as well as other 
materials, at low grain sizes, we consider the different types of atoms and 
local bonding geometries that exist in a nanocrystalline material. The main 
dramatic change that takes place in a polycrystalline material as its grain 
size enters the nm regime is the increase in the fraction of atoms residing 
near grain boundaries. Atoms at grain faces, edges or vertexes, as well as 
atoms near other discontinuities, will naturally form bonds that are weaker 
than those formed by atoms in the bulk. Such weaker bonds will then bend or 
stretch with greater ease, compared to the bonds in the crystalline region. 
This explains the softening of polycrystalline solids when the grain size is 
at the nanometer range. For much larger grain sizes, the number of 
grain-boundary atoms will be negligible compared to the number of bulk 
atoms; in this regime, the behavior of the material under mechanical load 
will be mostly determined by bulk defects, such as dislocations. 

To make this picture quantitative, let us divide the atoms in the 
polycrystalline material into three categories:

\begin{enumerate}
\item Atoms deep inside the grains, forming bonds that are similar to those in
the single-crystal material. The number of such atoms is proportional to
\textit{d}$^{3}$, where \textit{d} is the average grain size.
\item Atoms near the grain boundary; these behave similarly to surface or
interface atoms. The number of such atoms is proportional to \textit{d}$^{2}$.
\item Atoms near grain boundary edges; these are similar to kink surface
atoms, or atoms near dislocation cores. The number of such atoms is
proportional to \textit{d}.
\end{enumerate}
Of course, there will be other types of atoms, such as vertex atoms or atoms 
near topologic defects, but their number will be much smaller than the 
numbers of atoms falling in one of the aforementioned categories. The 
cohesive energy of the solid will be the sum of the energies of the three 
different atom types, multiplied by their respective numbers, and divided by 
the total number of atoms, which is proportional to \textit{d}$^{3}$. 
Therefore, the cohesive energy should be described by a function of the form

\begin{center}
\textit{E}$_{c}$ = \textit{E}$_{0}$ +\textit{a/d}+\textit{b/d}$^{2}$,
\end{center}

where \textit{a} and \textit{b} are constants, and \textit{E}$_{0}$ is the 
cohesive energy of the single crystal.

Indeed, such a function fits our data perfectly, the rms error being less 
than 0.5{\%}. Moreover, \textit{E}$_{0}$ is found to be -7.31 eV, very close 
to the calculated cohesive free energy of diamond at 300 K, which is -7.33 
eV. As B is proportional to the second derivative of the total energy with 
respect to the system volume, it can also be decomposed into contributions 
from bulk, interface and vertex atoms. As shown in Fig. 2, a quadratic 
function of 1/\textit{d} fits the results of the simulation very nicely. The 
constant value, 467 GPa, corresponding to the ideal monocrystalline solid, 
is only 5{\%} off the calculated value for diamond (see Table I). Such a 
decomposition of the total bulk modulus to a sum of atomic-level moduli has 
been used previously, in order to investigate the rigidity of amorphous 
carbon (Kelires 2000). 

A similar scaling law should also hold for the mass density of UNCD as a 
function of grain size, assuming that the volume per atom is different for 
atoms in grain boundaries and atoms in the bulk of grains. Fitting our data 
to a quadratic form of 1/\textit{d} gives \textit{$\rho 
$=}3.6-1.2/\textit{d}+0.4/\textit{d}$^{2}$. Again, the agreement of the 
constant value with the calculation for ideal diamond (Table I) is very good 
(3{\%}). 

Hardness is related to the electron density according to Gao et al.; the 
local electron density is proportional to the local mass density, as all C 
atoms have the same number of electrons. Therefore, hardness should also be 
decomposed into contributions from different kinds of atoms. As shown in 
Fig. 3, the hardness of UNCD can be fitted to a quadratic form of 
1/\textit{d. }Moreover, the constant term, showing the limit of hardness as 
\textit{d} goes to infinity, coincides with the hardness of diamond at 300 
K, calculated using the same method (see Table I).

The Young's and shear modulus of UNCD will not necessarily follow the same 
scaling law. As both moduli are related to bond bending, the nature of the 
inter-atomic bonds is perhaps equally important to their number. We tried to 
fit our simulation data to a quadratic form of 1/\textit{d}. Although the 
fit does not look disappointing, the rms error in the fits were 
significantly higher than those for the fits of \textit{E}$_{c}$, 
\textit{B}, \textit{$\rho $} or \textit{H}; moreover, the constant values 
deviate from the properties of diamond by more than 20{\%}. However, even 
such a poor agreement between model and simulation provides extra evidence 
that our model has some solid basis.

\section{Conclusions}

Using ultra-nanocrystalline diamond (UNCD) as a prototype for a 
polycrystalline covalent solid with grains at the nanometer region, we have 
observed softening of the material as the grain size decreases, in analogy 
with the reverse Hall-Petch effect observed in nanocrystalline metals. The 
effect is attributed to the increasing fraction of grain-boundary atoms as 
the grain size decreasing. A simple quadratic form in 1/\textit{d}, where 
\textit{d} is the average grain size, suffices to provide excellent fit of 
our results for cohesive energy, mass density, bulk modulus and estimated 
hardness, while it yields the correct values for bulk diamond. The measured 
Young's modulus of UNCD is reproduced well by the simulations. Our results 
provide further evidence that softening at low grain sizes might be a 
universal property for nanocrystalline solids.

\acknowledgments{The authors are grateful to Prof. Jacob Schi{\o}tz who shared
his programs for generating models of nanocrystalline solids, and acknowledge
inspiring discussions with Dr. Maria Fyta. This work was supported by a grant
from the Ministry of National Education and Religious Affairs in Greece
through the action ``$\mathrm{E}\Pi\mathrm{EAEK}$'' (programme ``$\Pi \Upsilon
\Theta$A$\Gamma$OPA$\Sigma$'').}

\end{document}